\documentstyle[12pt]{article}
\newlength{\aivwidth}   
\newlength{\tmpwidth}   

\newcommand{\phr}[1]{Phys.\ Rev.\ {\bf #1}}
\newcommand{\phrd}[1]{Phys.\ Rev.\ {\bf D#1}}
\newcommand{\phrl}[1]{Phys.\ Rev.\ Lett.\ {\bf #1}}
\newcommand{\nphb}[1]{Nucl.\ Phys.\ {\bf B#1}}
\newcommand{\phlb}[1]{Phys.\ Lett.\ {\bf B#1}}
\newcommand{\zphc}[1]{Z.\ Phys.\ {\bf C#1}}
\newcommand{\aph}[1]{Ann.\ Phys.\ {\bf #1}}
\newcommand{\phrp}[1]{Phys.\ Rep.\ {\bf #1}}
\newcommand{\ptph}[1]{Prog.\ Theor.\ Phys.\ {\bf #1}}

\newcounter{saveeqn}

\newcommand{\eqnew}{\setcounter{equation}{0}}

\newcommand{\be}{\begin{equation}}
\newcommand{\ee}{\end{equation}}
\newcommand{\bea}{\begin{eqnarray}}
\newcommand{\eea}{\end{eqnarray}}
\newcommand{\ba}{\begin{array}}
\newcommand{\ea}{\end{array}}
\newcommand{\eref}[1]{(\ref{#1})}
\newcommand{\nn}{\nonumber\\}

\newcommand{\D}{{\cal D}}
\newcommand{\dx}{d^4x\,}
\renewcommand{\L}[1]{{\cal L}_{#1}}
\newcommand{\lb}[1]{\bar{\cal L}_{#1}}
\renewcommand{\H}[1]{{\cal H}_{#1}}
\newcommand{\th}[1]{\tilde{\cal H}_{#1}}
\renewcommand{\d}{\partial}
\newcommand{\ph}{\varphi}
\newcommand{\ep}{\epsilon}
\newcommand{\df}{\delta^4(0)}
\newcommand{\da}{\dot{A}}
\renewcommand{\S}{\scriptstyle}
\newcommand{\Det}{{\rm Det}\,}
\newcommand{\tr}{{\rm tr}\,}
\newcommand{\dg}{\dagger}
\newcommand{\dph}{\dot{\ph}}
\newcommand{\tlis}{\tilde{\cal L}_I^S}
\newcommand{\ci}{\S F_{i0}^a\to\pi_i^a}
\newcommand{\cii}{\S A_0^a\to A_0^a-\lambda^a}
\newcommand{\ciii}{\S K_i^a=K_\lambda^a=0}
\newcommand{\civ}{\S \rho_i^a=\rho_\lambda^a=0}
\newcommand{\cv}{\S \rho_i^a=F_{i0}^a}
\newcommand{\cvi}{\S \rho_\lambda^a=0}
\newcommand{\cvii}{\S A_\mu\to U^\dg A_\mu
                      U-\frac{i}{g}U^\dg\d_\mu U}
\newcommand{\cviii}{\S \dph^a\to
                       Y^{-1}_{ab}\left(Y^{-1}_{cb}\pi_\ph^c-
                       MX_{bc}A^c_0\right)}
\newcommand{\cix}{\S A_0^a\to X_{ab} A_0^b+\frac{1}{M}Y_{ab}\dph^b}
\newcommand{\cx}{\S A_i^a\to X_{ab} A_i^b+\frac{1}{M}Y_{ab}\d_i\ph^b}
\newcommand{\cxi}{\S A_0^a \to \frac{1}{M}Y_{ba}^{-1}\pi_\ph^b}
\newcommand{\cxii}{\S \frac{v}{\sqrt{2}}U\to\Phi}
\newcommand{\cxiii}{\S \Phi \to \frac{v+h}{\sqrt{2}}U}
\newcommand{\cxiv}{\S \ph_a=0}
\newcommand{\cxv}{\S A_0^a\to\frac{1}{M}\pi_\ph^a}
\newcommand{\cxvi}{\S F_{\mu\nu}\to U^\dg F_{\mu\nu}U}
\newcommand{\cxvii}{\S F_{i0}^a\to X_{ab}F_{i0}^b}
\newcommand{\cxviii}{\S F_{ij}^a\to X_{ab}F_{ij}^b}
\newcommand{\cxix}{\S F_{i0}^a\to X_{ab}\pi_i^b}
\newcommand{\cxx}{\S \pi_\ph^a\to MY_{ba}A_0^b}
\title{Hamiltonian Quantization of Effective Lagrangians\\
with Massive Vector Fields}
\author{Carsten Grosse-Knetter\thanks{E-Mail:
knetter@physf.uni-bielefeld.de}
\\[5mm]Universit\"at Bielefeld\\Fakult\"at f\"ur Physik\\
W-4800 Bielefeld 1\\Germany}
\date{BI-TP 93/17\\ hep-ph/9304310 \\ April 1993}

\begin{document}
\begin{titlepage}
\thispagestyle{empty}
\maketitle
\thispagestyle{empty}
\begin{abstract}
Effective Lagrangians containing arbitrary
interactions of massive vector
fields are quantized within the Hamiltonian path
integral formalism. It is
proven that correct Hamiltonian quantization of these
models yields the same
result as naive Lagrangian quantization (Matthews's theorem).
This theorem holds for models without gauge freedom as well as
for (linearly or nonlinearly realized)
spontaneously broken gauge theories. The
Stueckelberg formalism, a procedure to rewrite
effective Lagrangians in a
gauge invariant way, is reformulated
within the Hamiltonian formalism
as a transition from a second class
constrained theory to an equivalent first class constrained theory.
The relations between linearly and nonlinearly
realized spontaneously broken
gauge theories are discussed. The quartically
divergent Higgs self interaction is derived
from the Hamiltonian path integral.
\end{abstract}
\end{titlepage}

\section{Introduction}
\typeout{Section 1}
\eqnew
Effective Lagrangians containing massive vector
fields with arbitrary
(non--Yang--Mills) self interactions have been
investigated very intensively in
the literature (see e.g. \cite{nongauge,nonlin,gauge}) in order
to parametrize possible deviations of
electroweak interactions from the standard model
with respect to experimental
tests of the $W^\pm$, $Z$ and $\gamma$ self couplings.
In \cite{nongauge,nonlin,gauge}
it is always implicitely assumed that
the Feynman rules can be directly
obtained from the effective Lagrangian,
i.e., the quadratic terms in the Lagrangian
yield the propagators and the
cubic, quartic, etc., terms yield the vertices. This simple
quantization rule is known as Matthews's theorem \cite{mat}.
Within the framework of the
the Feynman path intergral (PI) formalism
(where the Feynman rules follow from
the generating functional) it can be expressed as follows:
\begin{quote}{\it
Given a Lagrangian $\L{}$
with arbitrary interactions of massive
vector fields (among each other and with
other fields), the corresponding generating functional can
be written as a Lagrangian PI
\be Z[J]=\int\D\varphi\,\exp\left\{
i\int \dx [\L{quant}(\varphi,\d_\mu\varphi)
+J\varphi]\right\} \label{lpi}\ee
(where $\ph$ is a shorthand notation
for all fields in $\L{}$). If $\L{}$ has no gauge freedom,
the quantized
Lagrangian $\L{quant}$ occuring in the
PI is identical to the primordial one
\be \L{quant}=\L{}. \label{eq}\ee
If $\L{}$ has a gauge freedom, the generating functional\/
{\rm\eref{lpi}} is the same as the one obtained in
the Faddeev--Popov formalism\/
{\rm\cite{fapo}} with the quantized Lagrangian
\be \L{quant}=\L{}+\L{g.f.}+\L{FP\mbox{-}ghost}, \label{fp} \ee
which contains additional gauge fixing (g.f.)\ and ghost terms.}
\end{quote}
It is well known that,
in general, quantization has to be performed within
the Hamiltonian PI formalism.
The naive Lagrangian PI formalism, where \eref{lpi} with
\eref{eq} is taken as the ansatz
for the generating functional, can only
be directly applied to quantize
physical systems without derivative couplings and without
constraints. Thus, to prove
Matthews's theorem, one has
to derive the Lagrangian PI \eref{lpi} with
\eref{eq} or \eref{fp} within
the Hamiltonian PI formalism.

Matthews's theorem has been proven by Bernard and Duncan \cite{bedu}
for effective interactions of scalar fields; i.e.\
for models given by
nonsingular effective
Lagrangians. Massive vector fields, however, involve constraints.
Thus, one has to take into account the formalism of quantization
of constrained systems, which goes back to Dirac
\cite{dirac} and has been formulated within the
path integral formalism by
Faddeev \cite{fad} (for first class constrained, i.e.\
gauge invariant, systems)
and by Senjanovic \cite{sen} (for second class constrained, i.e.\
gauge noninvariant, systems).
Recent extensive treatises on this subject can be
found in \cite{gity,hete}.
In this paper, I will prove Matthews's theorem for
effective interactions of
massive vector fields taking into account this formalism.

Since it is in general not possible to find closed expressions for
the velocities
and the Hamiltonian in terms of the fields and the
generalized momenta within an
effective theory (if there are higher than second
powers of $\d_\mu\ph$ in the Lagrangian),
Bernard and Duncan assumed that
the effective interaction terms
are proportional to an $\epsilon$ with $\epsilon \ll
1$ and proved Matthews's theorem to a finite order in $\epsilon$.
I will proceed similarly;
I will assume that the vector boson self interactions are given by
Yang--Mills interactions (which can be treated
straightforwardly within the
Hamiltonian PI formalism
\cite{fad,sen,gity}) plus extra non--Yang--Mills interactions,
which are proportional to a small $\epsilon$. In the
following proof I will
only consider terms which are at most first
order in $\epsilon$, neglecting
higher powers of $\epsilon$. This treament is
justified when dealing with
phenomenologically motivated effective Lagrangians
\cite{nongauge,nonlin,gauge},
since these are considered to investigate the effects of
small deviations from the standard Yang--Mills couplings.

It will turn out that
the result \eref{eq} or \eref{fp} is only correct up to
additional quartically divergent terms%
\footnote{The $\df$ terms can be interpreted
as the contributions of the
loops of static ghost fields \cite{gkko}. Thus, they
do not contribute in the tree approximation.}, i.e.\ terms
proportional to $\delta^4(0)$. In \cite{bedu}
it is argued that these
terms can be neglected, since within dimensional
regularization $\delta^4(0)$ becomes zero. In fact, it is an open
question, how to interpret
divergences higher than logarithmic within an
effective (nonrenormalizable)
field theory \cite{bulo2}.
In this paper I will also neglect $\delta^4(0)$ terms when
establishing the equivalence
of Hamiltonian and Lagrangian quantization.
To give an example of such a term, I will derive the well known
quartically divergent Higgs
self-interaction term \cite{gkko,leya,lezj,giro}
from the Hamiltonian PI.

Recently, Lagrangians have been considered which contain
non--Yang--Mills self interactions
of massive vector fields within a
gauge invariant framework with spontaneously broken symmetry
\cite{nonlin,gauge,gkko,bulo}.
Thus, to justify the treatment
of these models within the (Lagrangian)
Faddeev--Popov formalism
\cite{fapo}, I will prove Matthews's theorem also for
spontaneously broken gauge theories (SBGTs).
To do this, I will first
consider SBGTs with a nonlinear
realization of the unphysical scalar fields.
Each of these models can be obtained by
applying a Stueckelberg transformation
\cite{stue} to a Lagrangian without gauge freedom \cite{gkko,bulo}
which is obtained by removing all unphysical scalar fields
from the gauge invariant Lagrangian and which
will be shown (within the Hamiltonian formalism)
to be the unitary gauge (U-gauge) of the original
SBGT. I will reformulate the Stueckelberg
formalism \cite{stue} within the
Hamiltonian formalism, thereby establishing the
equivalence of (nonlinear) gauge invariant Lagrangians and
the corresponding gauge noninvariant
Lagrangians. This
enables a generalization of Matthews's theorem to (nonlinearly
realized) SBGTs.

A priori it is
not clear that two Lagrangians related by a Stueckelberg
transformation are equivalent,
since such a transformation is not a simple
point transformation
because it involves derivatives of the unphysical scalar
fields; however, within the
Hamiltonian formalism this
equivalence can be properly shown. Within this
formalism no more ``Stueckelberg transformation''
is performed, instead, when
passing from the gauge
noninvariant (second class constrained) system to the
gauge invariant (first class constrained)
system, one enlarges the phase space
\cite{fash} by introducing new
(unphysical) variables and additional constraints
that express the new
variables in terms of the old ones.
Next, one uses the extra constraints to
rewrite the
Hamiltonian and the primordial constraints.
Then one half of the second class
constraints can be considered as first
class constraints and the other half as
gauge fixing conditions \cite{mira}.

The proof of Matthews's theorem for SBGTs goes then as follows:
Using the Stueckelberg formalism described above, I will show
that the generating funtional corresponding to a SBGT can be
written as a Lagrangian PI with the quantized Lagrangian being
identical to the U-gauge Lagrangian (i.e. the Lagrangian which is
obtained by removing all unphysical scalar fields from the gauge
invariant one). This generating funtional has been shown to be the
result of the FP procedure \cite{fapo} if the (U-gauge) g.f.\
conditions that all unphysical scalar fields become equal to zero
are imposed\footnote{\label{ug}
This is due to the fact that within
this special gauge there is no g.f. term (because the FP
$\delta$-function, which usually serves to introduce the g.f.\ term,
vanishes when performing the integration over the unphysical
scalar fields in order
to remove these fields from the Lagrangian)
and the ghost term can be expressed as a $\df$ term and thus be
neglected here \cite{gkko}.} \cite{gkko}. Then I will use the
equivalence of all gauges \cite{lezj,able}
in order to generalize the result
\eref{fp} to any other gauge.

Finally, I will prove Matthews's theorem
for Higgs models, i.e.\ for SBGTs
with linearly realized
scalar fields. Since each Higgs model is related to a
nonlinear Stueckelberg model by a simple point transformation
\cite{gkko,lezj,clt}, which becomes a
canonical transformation within
the Hamiltonian formalism and leaves the
Hamiltonian PI invariant, the result for
nonlinearly realized SBGTs can easily be
generalized to linearly realized SBGTs. As in the nonlinear case,
Matthews's theorem will first be derived for the special case of
the U-gauge and then be generalized to any other gauge.

My proof of Matthews's theorem will be restricted to
effective Lagrangians,
which do not depend on higher order
derivatives of the fields which depend on
first order derivatives of the vector fields only
through the non-Abelian field
strength tensor. (The latter requirement ensures
that the SBGTs corresponding
to such effective Lagrangians also do not involve higher order
derivatives.) This
includes the phenomenologically most important interactions
\cite{nongauge,nonlin,gauge}.

In this paper, I will only consider massive Yang--Mills
fields (of course with extra non--Yang--Mills interactions)
where all vector
bosons have equal masses and the corresponding SBGTs.
The results can easily
be generalized to any other effective Lagrangian with
massive vector bosons,
e.g.\ to electroweak models. In these cases the
treatment becomes formally
more complicated (in electroweak models there are extra
first class constraints
due to the unbroken subgroup and extra second class
constraints due to the
presence of fermions, which can, however, be treated
in a standard manner) but
the physically important features remain the same.
Thus, for clearness of
representation, I will restrict here to the investigation
of simple massive
Yang--Mills theories.

This paper is organized as follows: In section~2,
effective Lagrangians without gauge freedom
are quantized using the Hamiltonian PI formalism and
Matthews's theorem is proven for such models. In section~3,
the Stueckelberg
formalism is reformulated
within the Hamiltonian formalism, the equivalence of
an arbitrary effective theory
without gauge freedom and the corresponding
nonlinear SBGT
is established and Matthews's
theorem is proven for nonlinearly
realized SBGTs. In section~4, Higgs models
are considered and the above proof is extended to linearly realized
SBGTs. In section~5,
the quartically divergent Higgs self-interaction term
is derived from the Hamiltonian
PI. Section~6 is devoted to a summary of the results.

\section{Matthews's Theorem for Massive Vector Fields}
\typeout{Section 2}
\eqnew
In this this section, I will quantize a massive
Yang--Mills theory with
additional non--Yang--Mills interactions \cite{nongauge},
which are proportional to a parameter
$\ep$ (with $\ep{}\ll 1$), within the Hamiltonian PI formalism and
derive the simple Lagrangian form \eref{lpi}
with \eref{eq} of the generating functional
upon neglcting terms proportional to $\ep^2$ or to $\df$.

The effective Lagrangian has the form
\be\L{}= \L{0}+\ep\L{I}=
-\frac{1}{4}F^{\mu\nu}_aF_{\mu\nu}^a+\frac{1}{2}M^2A^\mu_aA_\mu^a
+\ep\L{I}(A_\mu^a,F_{\mu\nu}^a)\label{leff}\ee
($a=1,\ldots ,N$) with
\be F_{\mu\nu}^a=\d_\mu A_\nu^a-\d_\nu A_\mu^a +g f_{abc}A_\mu^b
A_\nu^c.\label{fst}\ee
For the non--Yang--Mills part of the effective interactions,
given by
$\L{I}$, I make the following assumptions:
\begin{itemize}\label{req}
\item $\L{I}$ does not depend on higher order derivatives.
\item $\L{I}$ depends on first order
derivatives of $A_\mu^a$ only through the
non-Abelian field strength tensor\footnote{I will
need this requirement only
in the next two sections to investigate the
SBGTs corresponding to $\L{}$. For the
treatment of this section, the weaker requirement
that $\L{I}$ does not
depend on $\dot{A}^a_0$ is sufficient.} $F_{\mu\nu}^a$ \eref{fst}.
\end{itemize}
These conditions are fulfilled by
the phenomenologically most important
effective interactions, especially by
all nonstandard $P$, $C$ and $CP$
invariant trilinear interactions
of electroweak vector bosons \cite{nongauge}.

{}From \eref{leff} one finds the momenta
\bea \pi_0^a&=&\frac{\d\L{}}{\d \da^0_a}=0,\label{pn}\\
\pi_i^a&=&\frac{\d\L{}}{\d \da^i_a}=F_{i0}^a+\ep
\frac{\d\L{I}}{\d \da^i_a}
=\dot{A}^i_a+\d_iA^a_0+gf_{abc}A^b_iA^c_0+\ep
\frac{\d\L{I}}{\d \da^i_a}.
\label{pi}\eea
\eref{pi} can be solved for $\da_a^i$ (to first order in $\ep$)
\be \da^i_a=\left.\pi_i^a-\d_i A_0^a-gf_{abc}A^b_iA^c_0-
\ep\frac{\d\L{I}}{\d \da^i_a}\right|_{\ci}+O(\ep^2). \label{dai}\ee
The Hamiltonian is given by
\bea \H{}&=&\pi_\mu^a\da^\mu_a-\L{}\nn
&=&\frac{1}{2}\pi_i^a\pi_i^a-\pi_i^a\d_iA^a_0-gf_{abc}\pi^a_i
A^b_iA^c_0+\frac{1}{4}F_{ij}^aF_{ij}^a-\frac{1}{2}M^2 (A_0^aA_0^a-
A_i^aA_i^a)\nn
&&-\ep\lb{I}+O(\ep^2),\label{heff}\eea
where $\lb{I}$ is defined as
\be \lb{I}\equiv\L{I}|_{\ci}.\label{lib} \ee
\eref{pn} yields the primary constraints
\be\phi_1^a=\pi_0^a=0.\label{pc}\ee
The secondary constraints are obtained from the
requirement that the primary
constraints must be consistent
with the equations of motion, i.e.\ the relations
\be \dot{\phi}_1^a=\{\phi_1^a,H\}=0\label{eom}\ee
must be fulfilled. This yields
\be \phi_2^a=\d_i\pi_i^a-gf_{abc}\pi^b_i A^c_i-
M^2 A_0^a-\ep\frac{\d\lb{I}}
{\d A_0^a} +O(\ep^2)=0.\label{sc}\ee
There are no further constraints. The Poisson
brackets of the primary and the
secondary constraints are
\be\{\phi_1^a(x),\phi_2^b(y)\}=
\left(M^2\delta^{ab}+\ep\frac{\d\lb{I}}{\d A_0^a
\d  A_0^b} +O(\ep^2)\right)\delta^4(x-y).\label{pb}\ee
Since $\{\phi_1^a(x),\phi_1^b(y)\}=0$, one finds
\be \Det^{\frac{1}{2}}\{\Phi^a,\Phi^b\}=(-1)^{N+1}
\Det\{\phi^a_1,\phi^b_2\}\ne 0 \label{det}\ee
(with $\Phi=(\phi_1,\phi_2)$). Thus, the
constraints are second class. This is
due to the fact that $\L{}$ is gauge
noninvariant, since the mass term and
(in general) the non--Yang--Mills
interactions in $\L{I}$ break gauge invariance
explicitely.

The generating functional for a second class constrained system is
generally given by \cite{sen,gity,hete}
\be Z[J]=\int\prod_{\mu,a}
\D A_\mu^a\D\pi_\mu^a\,\exp\left\{i\int\dx[\pi_\mu^a\da^\mu_a-\H{}+
J_\mu^aA^\mu_a]\right\}\prod_a(\delta(\phi_1^a)\delta(\phi_2^a))
\Det^{\frac{1}{2}}
\{\Phi^a,\Phi^b\}.\label{hpi}\ee
The determinant in \eref{hpi} only yields $\df$ terms which are
neglected here. This can easily be seen from
\eref{pb} and \eref{det}
by use of the
identity\footnote{\label{foot}Another
way to see this is to rewrite the determinant as a
functional integral over Grassmann variables
which yields a ghost term
$\L{ghost}=-M^2\eta_a^\ast\eta_a-
\ep\eta^\ast_a\frac{\d\lb{I}}{\d A_0^a \d
A_0^b}\eta_b+O(\ep^2)$. The ghost fields
are static, i.e.\ there are no kinetic terms for
them, only mass terms and couplings to the $A^\mu_a$ fields.
This means, all
ghost propagators are simply inverse masses and thus
all ghost loops are
quartically
divergent.  Thus, the ghost term can be replaced by a
$\df$ term which
yields the same contribution to matrix elements as the ghost loops
\cite{gkko}.} \cite{sast}
\be \Det(M_{ab}(x)\delta^4(x-y))=\exp\left\{\delta^4(0)
\int\dx\ln({\rm
det}\,M_{ab}(x))\right\}\label{expdet}\ee
(where ``Det'' expresses the functional determinant
and ``det'' the ordinary one).
Dropping the determinant, integrating
out the $\pi_0^a$ due to the presence of
$\prod_a\delta(\pi_0^a)$ in \eref{hpi}, and using the relation
\be \prod_a\delta(\phi_2^a)\propto\int\prod_a\D\lambda^a\,
\exp\left\{-i\int\dx\lambda^a\phi_2^a\right\}\label{exp}\ee
one finds
\bea Z[J]&=&\int\prod_{\mu,a}\D A_\mu^a\prod_{i,a}
\D\pi_i^a\prod_a\D\lambda^a\,
\exp\Bigg\{ i\int\dx\Bigg[
-\frac{1}{2}\pi_i^a\pi_i^a\nn&&+  \pi_i^a
(\da^i_a+\d_i(A_0^a+\lambda^a)+
gf_{abc}A_i^b(A_0^c+\lambda^c))-\frac{1}{4}F_{ij}^a
F_{ij}^a\nn&&+\frac{1}{2}M^2((A_0^a+\lambda^a)
(A_0^a+\lambda^a)-\lambda^a
\lambda^a
-A_i^aA_i^a)\nn&&+\ep\left(\lb{I}+\lambda^a
\frac{\d\lb{I}}{\d A_0^a}\right)
+O(\ep^2)+J_\mu^aA^\mu_a\Bigg]\Bigg\}.\eea
The substitution\footnote{After this substitution the source
$J_0^a$ becomes coupled to  $A_0^a-\lambda^a$ instead of $A_0^a$.
However it
does not affect physical matrix elements to
remove the coupling of $\lambda^a$
to $J_0^a$ \cite{able}.}
\be A_0^a\to A_0^a-\lambda^a,\label{subs}\ee
which obviously leaves the functional integration measure
invariant, yields
\bea Z[J]&=&\int\prod_{\mu,a}\D A_\mu^a\prod_{i,a}\D\pi_i^a
\prod_a\D\lambda^a\,
\exp\bigg\{ i\int\dx\bigg[
-\frac{1}{2}\pi_i^a\pi_i^a+\pi_i^a F^a_{i0}-
\frac{1}{4}F_{ij}^a
F_{ij}^a\nn&&+\frac{1}{2}M^2( A_0^aA_0^a-\lambda^a\lambda^a
-A_i^aA_i^a)-\th{I}(A_\mu^a,\d_iA_\mu^a,\pi_i^a,\lambda^a)
+J_\mu^aA^\mu_a\bigg]\bigg\}.\label{step1}\eea
with
\be \th{I}(A_\mu^a,\d_iA_\mu^a,\pi_i^a,\lambda^a)
\equiv-\ep\left(\lb{I}+\lambda^a
\frac{\d\lb{I}}{\d A_0^a}\right)\Bigg|_{\cii}+O(\ep^2).
\label{thi}\ee

Now the procedure of Bernard and Duncan \cite{bedu}
can be generalized
to the model considered here.
Introducing sources $K_i^a$ and $K_\lambda^a$ coupled to $\pi_i^a$
and $\lambda^a$, one can rewrite \eref{step1} as
\bea \!\!\!
Z[J]&=&\int\prod_{\mu,a}\D A_\mu^a\,\exp\left\{-i\int\dx\th{I}\left
(A_\mu^a,\d_iA_\mu^a,
\frac{\delta}{i\delta K_i^a},\frac{\delta}{i\delta
K_\lambda^a}\right)\right\}
\nn&&\times\int\prod_{i,a}\D\pi_i^a\prod_a\D\lambda^a\,
\exp\bigg\{i\int\dx\bigg[
-\frac{1}{2}\pi_i^a\pi_i^a+  \pi_i^aF_{i0}^a-\frac{1}{4}F_{ij}^a
F_{ij}^a\nn &&\quad+\frac{1}{2}M^2( A_0^aA_0^a-\lambda^a\lambda^a
-A_i^aA_i^a)+J_\mu^aA^\mu_a+\pi_i
^aK_i^a+\lambda^aK^a_\lambda)\bigg]\bigg\}\bigg|_{\ciii}.\eea
Performing the Gaussian intergrations over $\pi_i^a$ and
$\lambda^a$ one gets
\bea Z[J]&=&\int\prod_{\mu,a}\D A_\mu^a\,\exp\left\{
i\int\dx[\L{0}+J_\mu^aA^\mu_a
]\right\}\nn&&\times\exp\left\{-i\int\dx\th{I}
\left(A_\mu^a,\d_iA_\mu^a,
\frac{\delta}{i\delta K_i^a},\frac{\delta}
{i\delta K_\lambda^a}\right)\right\}
\nn&&\times\exp\left\{
i\int\dx\left[\frac{1}{2}K_i^aK_i^a+\frac{1}{2}K_\lambda^aK_
\lambda^a
+K_i^aF_{i0}^a\right]\right\}\bigg|_{\ciii}\eea
where $\L{0}$ is the massive Yang--Mills part of the
effective Lagrangian
\eref{leff}. The use of the functional identity \cite{col}
\be F\left[\frac{\delta}{i\delta K}\right]G[K]\Bigg|_{K=0}=
G\left[\frac{\delta}{i\delta \rho}\right]F[\rho]\Bigg|_{\rho=0}\ee
yields
\bea Z[J]&=&\int\prod_{\mu,a}\D A_\mu^a\,\exp\left\{
i\int\dx[\L{0}+J_\mu^aA^\mu_a
]\right\}\nn&&\times\exp\left\{\int\dx\left[{}-\frac{i}{2}
\sum_{i,a}\left(\frac{\delta}
{\delta\rho_i^a}\right)^2-\frac{i}{2}
\sum_a\left(\frac{\delta}
{\delta\rho_\lambda^a}\right)^2+F^a_{i0}
\left(\frac{\delta}{\delta\rho_i^a}\right)\right]\right\}\nn&&\times
\exp\left\{-i\int\dx\th{I}\left (A_\mu^a,\d_iA_\mu^a,
\rho_i^a,\rho_\lambda^a\right)\right\}\bigg|_{\civ}.\label{form}\eea
Since $\th{I}$ \eref{thi} is proportional to
$\ep$, the third exponetial in \eref{form}
can be expanded in powers of $\ep$:
\be \exp\left\{-i\int\dx\th{I}\left (A_\mu^a,\d_iA_\mu^a,
\rho_i^a,\rho_\lambda^a\right)\right\}=1-i\int\dx\th{I}\left
(A_\mu^a,\d_iA_\mu^a,\rho_i^a,\rho_\lambda^a\right)+O(\ep^2).
\label{hexp}\ee
Obviously, second order
functional derivatives with respect to the $\rho$'s
acting on this expression yield terms which are proportional to
$\ep^2$ or to $\df$ and
which are both neglected here. Thus,
the second order derivatives in the second
exponential in \eref{form} can be omitted.
The second and the
third exponential in \eref{form} together reduce to
\bea &&\exp\left\{\int\dx F^a_{i0}
\left(\frac{\delta}{\delta\rho_i^a}\right)\right\}
\exp\left\{-i\int\dx\th{I}\left (A_\mu^a,\d_iA_\mu^a,
\rho_i^a,\rho_\lambda^a\right)\right\}\Bigg|_{\civ}\nn&&
\qquad\qquad\qquad=
\exp\left\{-i\int\dx\th{I}\left (A_\mu^a,\d_iA_\mu^a,
\rho_i^a,\rho_\lambda^a\right)\right\}\bigg|_{\ba{l}\cv\\
\cvi \ea}.\label{tl}
\eea
With the definitions of $\th{i}$ \eref{thi} and of $\lb{I}$
\eref{lib}
one finds
\be \th{I}\left (A_\mu^a,\d_iA_\mu^a,
\rho_i^a,\rho_\lambda^a\right)\bigg|_{\ba{l}\cv\\ \cvi \ea} =- \ep
\L{I}(A_a^\mu, F^{\mu\nu}_a) +O(\ep^2)\label{simp}.\ee
The insertion of \eref{tl} with \eref{simp} into \eref{form} yields
(apart from $\ep^2$ and $\df$ terms)
\be Z[J]=\int\prod_{\mu,a}\D A_\mu^a\,\exp\left\{i\int\dx[\L{0}+
\ep\L{I}+
J_\mu^aA^\mu_a]\right\},\label{result}\ee
which is the expected result, namely
the naive Lagrangian path integral \eref{lpi} with \eref{eq}.
Thus, Matthews's theorem is proven for effective self
interactions of massive
vector fields (within a gauge noninvariant framework).

This proof can easily be generalized to Lagrangians which
also contain effective interactions of the massive vector
fields with other
fields (scalar, fermion or additional vector fields). To
derive this result,
one adds in \eref{leff}
the kinetic and mass terms of the extra fields as well as the
couplings without
derivatives to $\L{0}$ and the
derivative couplings to $\L{I}$
and then goes through the same procedure as above.
Thus, Matthews's
theorem also holds for effective vector--fermion and vector--scalar
interactions\footnote{An application of this result,
which will become important
in section 4, is to consider $\L{0}$ as the U-gauge
Lagrangian of a (minimal) Higgs model,
while $\L{I}$ contains additional effective interactions
of the vector
and Higgs fields.}.

\section{The Stueckelberg Formalism}
\typeout{Section 3}
\eqnew
In this section, I will generalize Matthews's theorem to
SBGTs with nonlinearly
realized symmetry which contain arbitrary gauge boson self
interactions within
a gauge invariant framework \cite{nonlin,gkko,bulo}. It has been
shown in \cite{gkko,bulo} that each theory given by an effective
Lagrangian of the type \eref{leff} can be rewritten as a
(nonlinearly realized)
SBGT by applying Stueckelberg transformations \cite{stue}.
On the other hand,
each nonlinear SBGT (without higher derivatives)
can be obtained by applying a Stueckelberg transformation to
a Lagrangian of type \eref{leff}. Thus, I will
reformulate the Stueckelberg formalism within the Hamiltonian
formalism in order to show the equivalence of effective
Lagrangians which are related by Stueckelberg transformations.

The Stueckelberg formalism can be most easily formulated within the
matrix notation. With $t_a$ being the generators of the gauge
group, which are
orthonormalized due to
\be \tr(t_a,t_b)=\frac{1}{2}\delta_{ab} \label{tr}\ee
one defines
\bea A_\mu&\equiv&A_\mu^a t_a,\label{amu}\\
\ph&\equiv&i\frac{g}{M}\ph^at_a,\\
U&\equiv&\exp\ph.\eea
The $\ph^a$ are the unphysical pseudo-Goldstone scalars.
The Stueckelberg
transformation is defined as:
\be A_\mu\to -\frac{i}{g}U^\dg D_\mu U=
U^\dg A_\mu U-\frac{i}{g}U^\dg\d_\mu U
=U^\dg A_\mu U+\frac{1}{M}\d_\mu\ph^{a}U^\dg Q_a\label{stue}\ee
($D_\mu U$ is the covariant derivative of $U$) with
\be Q_a\equiv \left(t_a +(\ph t_a+t_a\ph) +\frac{1}{2!}(\ph^2t_a+\ph
t_a\ph+t_a\ph^2)+\ldots \right). \ee
The Stueckelberg transformation \eref{stue} {\em formally}
acts like a gauge
transformation, however, with the gauge parameters being
replaced by the
pseudo-Goldstone fields. Thus, it effects only the mass term
and the effective
interaction term $\L{I}$ in \eref{leff} but not the gauge
invariant Yang--Mills
term $-\frac{1}{4}F^{\mu\nu}_aF_{\mu\nu}^a$.
\eref{stue} can be written in components by
multiplying with
$2t_a$ and taking the trace. With \eref{tr} and \eref{amu} one finds
\be A_\mu^a\to X_{ab} A_\mu^b+\frac{1}{M}Y_{ab}\d_\mu\ph^b
\label{st}\ee
where the matrices $X$ and $Y$ are defined as
\bea X_{ab}&\equiv& 2\tr(U^\dg t_bUt_a),\label{x}\\
Y_{ab}&\equiv&2\tr(U^\dg Q_b t_a).\label{y}\eea
$X$ and $Y$ are nonpolynomial expressions in the
pseudo-Goldstone fields
$\ph^a$. They do not depend on
the derivatives $\d_\mu \ph^a$ and, due to
\eref{tr}, they become unity
matrices for vanishing $\ph^a$:
\be X_{ab}(\ph^a=0)=Y_{ab}(\ph^a=0)=\delta_{ab} \label{van}.\ee

The SBGT corresponding to the effective Lagrangian \eref{leff}
is
\be \L{}^S\equiv\L{}\big|_{\cvii}.\label{leffs}\ee
$\L{}$ can be recovered from $\L{S}$ simply by removing all
unphysical scalar
fields in $\L{}^S$
\be \L{}=\L{}^S\big|_{\ph_a=0}. \ee
The non--Yang--Mills
part of the effective interactions is given by the gauge
invariant term
$\L{I}^S$, which is obtained by applying $\eref{stue}$ to $\L{I}$.
$\L{}^S$ describes a generalized gauged nonlinear $\sigma$-model
with extra non-Yang--Mills vector boson self interactions
\cite{gkko}.
Each nonlinearly realized effective SBGT (without higher
derivatives) given by
a Lagrangian $\L{}^S$ can be constructed by
applying \eref{stue} to an
effective Lagrangian $\L{}$ \eref{leff}, which is obtained
by removing the
pseudo-Goldstone fields
in $\L{}^S$. I will prove that the Lagrangians
$\L{}$ and $\L{}^S$ describe equivalent physical
systems\footnote{For simple gauged nonlinear
$\sigma$-models
without effective interactions this equivalence has been shown
within the
Hamiltonian formalism in \cite{pape}.}.
This is not obvious because the
Stueckelberg transformation \eref{stue} involves derivatives of the
pseudo-Goldstone fields and from the Lagrangian point of view
one can only
argue that two Lagrangians which are related by a point
transformation (i.e. a transformation
which does not involve derivatives)
are equivalent. I will show within the Hamiltonian
formalism that $\L{}$ is the U-gauge of $\L{}^S$,
i.e., the U-gauge of a nonlinear effective SBGT is simply
obtained by dropping
all unphysical scalar fields (as one naively expects).

One can easily see that, if $\L{}$ satisfies the conditions listed
at the beginning of section~2,
$\L{S}$ also fulfils these requirements, since the
field strength tensor $F_{\mu\nu}=F^a_{\mu\nu}t_a$ transforms under
Stueckelberg transformations due to
\be F_{\mu\nu}\to U^\dg F_{\mu\nu}U \label{fs}\ee
or, written in components,
\be F_{\mu\nu}^a\to X_{ab}F_{\mu\nu}^b.\ee
For the subsequent treatment it is convenient to rewrite
\eref{leffs} as
\be \L{}^S=\L{}\bigg|_{\ba{l}\cvii\\ \cxvi\ea}=
\L{}\bigg|_{\ba{l} \cix\\\cx\\
\cxvii\\ \cxviii\ea}, \label{ls}\ee
where the following convention has been used: While
in \eref{leffs} the
Stueckelberg transformation is applied to $A_\mu$
{\em everywhere\/} in $\L{}$
(which automatically implies
the transformation of $F_{\mu\nu}$ \eref{fs})
it is in \eref{ls} only applied to the $A_\mu$ field
where it does not occur as a part of
the field strength tensor $F_{\mu\nu}$, and $F_{\mu\nu}$
becomes then transformed
seperately. I will use this convention throughout this section.

The momenta conjugate to the fields in $\L{}^S$ are
\bea \pi_0^a&=&\frac{\d\L{}^S}{\d \da^0_a}=0,\label{pns}\\
\pi_i^a&=&\frac{\d\L{}^S}{\d \da^i_a}=F_{i0}^a+\ep\frac{\d\L{I}^S}
{\d \da^i_a}
=\dot{A}^i_a+\d_iA^a_0+gf_{abc}A^b_iA^c_0+\ep
\frac{\d\L{I}^S}{\d \da^i_a},
\label{pis}\\
\pi_\ph^a&=&\frac{\d\L{}^S}{\d\dph^a}=MY_{ca}\left(X_{cb}A_0^b
+\frac{1}{M}Y_{cb}\dot{\ph}^b\right)+\ep\frac{\d\L{I}^S}{\d\dph^a}.
\label{pps}\eea
To first order in $\ep$ one finds the velocities
\bea \!\!\!\da^i_a&=&\left.\pi_i^a-\d_i A_0^a-gf_{abc}A^b_iA^c_0-
\ep\frac{\d\L{I}^S}{\d \da^i_a}\right|_{\ba{l}\ci\\ \cviii\ea}+
O(\ep^2),\\
\!\!\!\dph^a&=&Y^{-1}_{ab}\left(
Y^{-1}_{cb}\left(\pi_\ph^c-\left.\ep\frac{\d\L{I}^S}{\d\dph^c}
\right|_{\ba{l}\ci\\ \cviii\ea}\right)-MX_{bc}A^c_0\right)+
O(\ep^2)\eea
and the Hamiltonian
\bea
\!\H{}^S&=&\pi_\mu^a\da^\mu_a+\pi_\ph^a\dph^a-\L{}^S
\nn&=&\frac{1}{2}\pi_i^a\pi_i^a-\pi_i^a\d_iA^a_0-gf_{abc}\pi^a_i
A^b_iA^c_0+\frac{1}{4}F_{ij}^aF_{ij}^a\nn&&-\frac{1}{2}M^2
\left(\sum_a\left(X_{ab}A_0^b\right)^2-\sum_a\left(X_{ab}A_i^b+
\frac{1}{M}Y_{ab}\d_i\ph^b\right)^2\right)\nn&&+\frac{1}{2}\sum_a
(Y^{-1}_{ba}\pi_\ph^b-M X_{ab}A_0^b)^2
-\ep\lb{I}^S+O(\ep^2).\label{hs}\eea
$\lb{I}^S$ is defined as
\be \lb{I}^S\equiv\L{I}^S\bigg|_{\ba{l}\ci\\ \cviii\ea} =
\L{I}\Bigg|_{\ba{l}\cxi \\ \cx\\ \cxix \\\cxviii\ea} \label{con}\ee
(where \eref{ls} has been used).
The primary constraints are
\be\phi_1^a=\pi_0^a=0\label{pcs}\ee
and the secondary constraints, obtained analogously to
\eref{eom}, are
\be \phi^a_2=\d_i\pi_i^a-gf_{abc}\pi_i^bA_i^c
-MX_{ba}Y^{-1}_{cb}\pi^c_\ph =0.\label{scs}\ee
There are no terms proportional to $\ep$ in \eref{scs}, since,
due to \eref{con},
$\lb{I}^S$ does not depend on $A_0^a$ (neither directly nor through
$F_{i0}^a$)\footnote{In fact, to all orders in
$\ep$, $A_0^a$ becomes replaced by
$\frac{1}{M}Y_{ba}^{-1}\pi_\ph^b$ and
$F_{i0}^a$ by $X_{ab}\pi_i^b$. Thus,
$\H{I}^S$ does not depend on $A_0^a$ and \eref{scs} holds exactly.}.
There are no further constraints.
The constraints are first class due to the gauge freedom of $\L{}^S$
\eref{leffs}.

Since in first class constrained systems the solutions of
the equations of
motion contain undetermined Lagrange
multipliers, one has to remove this
ambiguity by imposing additional
gauge fixing (g.f.)\ conditions \cite{fad,gity,hete},
such that the number of g.f.\ conditions is
equal to the number of first class constraints. Constraints
and g.f.\
conditions together have to form a system of second
class constraints being
consistent with the equations of motion. A convenient
way to construct these conditions \cite{gity} is to start with
primary
g.f.\ conditions
$\chi_1^a$ and to construct secondary g.f.\ conditions by
demanding
\be \{\chi_1^a,H^S\}=0\label{congf}\ee
which ensures consistency with the equations of motion.
To prove the equivalence of $\L{}$ \eref{leff} and $\L{}^S$
\eref{leffs} it is
most convenient to construct the U-gauge by imposing
the primary g.f.\ conditions
\be \chi_1^a=\ph^a =0.\label{pgf}\ee
\eref{congf} yields then the
secondary g.f.\ conditions\footnote{The g.f.\
conditions do not fulfil
Faddeev's requirement $\{\chi^a,\chi^b\}=0$
\cite{fad}. In fact, this
restriction is unnecessary \cite{gity,hete,kadi}.}
\bea\chi_2^a&=&Y^{-1}_{ab}\left(
Y^{-1}_{cb}\pi^c_\ph-MX_{bc}A_0^c\right)-
\ep\frac{\d\lb{I}^S}{\d\pi_\ph^a}+O(\ep^2)=\nn
&=&Y_{ab}^{-1}\left(Y^{-1}_{cb}\pi^c_\ph-MX_{bc}A_0^c\right)-
\ep \frac{1}{M}Y^{-1}_{ab}\frac{\d\tlis}{\d A_0^b}+O(\ep^2)=0
\label{sgf}\eea
with the definition
\be\tlis\equiv\lb{I}^S\bigg|_{\cxx}=
\L{I}\Bigg|_{\ba{l}\cx\\\cxix\\\cxviii\ea}.
\label{tlis}\ee
Using the primary g.f.\
conditions\footnote{The insertion of the g.f.\
conditions into the Hamiltonian
and the other constraints corresponds
de facto to a redefinition of the Lagrange
multipliers in the total Hamiltonian,
i.e.\ the Hamiltonian from which follow
the equations of motion:
$\H{T}^S=\H{}^S+\lambda_a\Phi_a+
\tilde{\lambda}_a\chi_a$ (where $\Phi_a$ stands
for all constraints, $\chi_a$
for all g.f.\ conditions and the
$\lambda_a$ and $\tilde{\lambda}_a$ are the Lagrange multipliers.)}
\eref{pgf},  the relation
\eref{van} and the defintions of $\tlis$ \eref{tlis},
$\lb{I}^S$ \eref{con}
and $\lb{I}$ \eref{lib} one can
express the Hamiltonian \eref{hs}, the
secondary constraints \eref{scs} and g.f.\ conditions \eref{sgf} as
\bea\H{}^S&=&\frac{1}{2}\pi_i^a
\pi_i^a-\pi_i^a\d_iA^a_0-gf_{abc}\pi^a_i
A^b_iA^c_0+\frac{1}{4}F_{ij}^aF_{ij}^a
\nn&& -\frac{1}{2}M^2(A_0^aA_0^a-A_i^a
A_i^a)+\frac{1}{2}\sum_a(\pi_\ph^a-MA_0^a)^2-\ep\lb{I}\Big|_{\cxv}
+O(\ep^2),\label{hs2}\\
\phi^a_2&=&\d_i\pi_i^a-gf_{abc}\pi_i^bA_i^c
-M\pi^a_\ph+O(\ep^2) =0,\label{scs2}\\\chi_2^a&=&\pi^a_\ph-MA_0^a-
\ep \frac{1}{M}\frac{\d\lb{I}}{\d A_0^a}=0.\label{sgf2}\eea
Applying the secondary g.f.\ condition \eref{sgf2}, one can
rewrite the
Hamiltonian  \eref{hs2} as \eref{heff} and the constraints
\eref{pcs}, \eref{scs2} as \eref{pc}, \eref{sc} (to first
order in $\ep$),
i.e. as the Hamiltonian and
the constraints corresponding
to the gauge noninvariant Lagrangian $\L{}$
\eref{leff}.
Finally, the g.f.\ conditions \eref{pgf}
and \eref{sgf2} can be
omitted, since they involve the fields $\ph^a$ and $\pi_\ph^a$ and
neither the Hamiltonian nor the constraints depend on
these fields anymore.
Thus, the Lagrangians $\L{}$ and $\L{}^S$ in \eref{leffs}
describe equivalent physical
sytems, $\L{}$ being the U-gauge of $\L{}^S$.

Due to this equivalence one can quantize $\L{}^S$ as described
in the previous
section; the generating funtional turns
out to be \eref{result}. This, however,
is identical (apart from $\df$ terms, which are
neglected here) to the generating functional obtained in the
(Lagrangian)
Faddeev--Popov formalism \cite{fapo} if one imposes the (U-gauge)
g.f.\ conditions \eref{pgf}\footnote{Remember footnote \ref{ug}.}
\cite{gkko}. Due to the equivalence of all gauges
\cite{lezj,able},
\eref{result} yields the same $S$-matrix elements
as the Faddeev--Popov
PI in any other
gauge (e.g.\ $\rm R_\xi$
gauge, Lorentz gauge, Coulomb gauge) given by \eref{lpi}
with \eref{fp}. Thus, Matthews's theorem also holds for
SBGTs with nonlinearly
realized symmetry.

The above procedure shows how to interpret the
Stueckelberg formalism on the
Hamiltonian level. While the gauge noninvariant
Lagrangian $\L{}$ is related
to the gauge invariant Lagrangian $\L{}^S$ by a Stueckelberg
transformation \eref{stue}, one can pass from the
second class constrained
Hamiltonian $\H{}$ to the first class constrained
Hamiltonian $\H{}^S$ by the
following procedure: One enlarges the phase space by
introducing the unphysical
variables $\ph^a$ and $\pi_\ph^a$ and the extra constraints
$\eref{pgf}$ and
\eref{sgf}, which make the new variables dependent on the
others and leaves
the number of physical degrees of freedom unchanged.
Next, one rewrites,
using the additional constraints \eref{pgf} and \eref{sgf},
the Hamiltonian as \eref{hs} and
the primordial constraints as \eref{pcs} and
\eref{scs}. Finally, half of the constraints, namely the new
ones, are
considered as g.f.\ conditions\footnote{A similar transition
from a second
class constrained system to a first class
constrained system has recently been investigated in several
works \cite{mira}.
However, there no phase space enlargement is performed
with the outcome, that the
resulting model contains only half as much first class constraints
as the original model has second class constraints.
In my treatment the number of constraints remains
unchanged, since, due to the phase space enlargement,
new constraints are
introduced.
The method of connecting first and second class constrained systems
by performing a phase space enlargement goes back to \cite{fash}.}.

\section{Higgs Models}
\typeout{Section 4}
\eqnew
Finally, Matthews's theorem has to be proven for SBGTs with
linearly realized
symmetry, i.e.\ Higgs models,
which contain effective (non--Yang--Mills) gauge boson
self interactions
\cite{gauge,gkko}. This result will simply be obtained
by showing the
equivalence of a  linear Higgs model to a nonlinear
Stueckelberg model (with
(an) additional physical scalar(s)).

Since the Higgs model
corresponding to a massive Yang--Mills theory
cannot be written in a general form for an
arbitrary gauge group, I
restrict to the case of SU(2) symmetry
(i.e. $t_a =\frac{1}{2}\tau_a$,
$a=1,2,3$). The extension to other gauge groups is straightforward.

Any effective Lagrangian \eref{leff} can be extended to a
Higgs model by constructing the Stueckelberg Lagrangian
\eref{leffs} and then
introducing a physical scalar field
$h$ and linearizing the scalar sector of the theory \cite{gkko} via
\be \frac{v}{\sqrt{2}}U\equiv\frac{v}{\sqrt{2}}
\exp\left(\frac{i\ph_a\tau_a}{v}\right)\quad\to\quad
\Phi\equiv\frac{1}{\sqrt{2}} ((v+h){\rm \bf 1} +
i\ph_a\tau_a)\label{phi}\ee
where $v$ is the vacuum expectation value
of the Higgs field, $v=\frac{2M}{g}$.
The Lagrangian of the Higgs model corresponding to \eref{leff}
becomes
\be \L{}^H=\L{}^S\big|_{\cxii}-V(\Phi)\label{leffh}\ee
with the Higgs self interaction potential $V(\Phi)$
which yields the nonvanishing vacuum expectation value.
In distinction from $\L{}^S$ \eref{leffs}, $\L{}^H$ is is {\em
not} equivalent to the effective
Lagrangian $\L{}$, since there is an additional
physical degree of freedom. However, $\L{}^H$ contains the
same effective
vector boson self interactions as $\L{}$ and $\L{}^S$.
In fact, $\L{}^S$
is the
limit of $\L{}^H$ for infinite Higgs mass \cite{nonlin,gkko}.
Each effective Higgs model (without higher derivatives) can
be constructed this way from a Lagrangian of type \eref{leff}.

To extend the results of the previous two sections
to the Lagrangian
\eref{leffh}, one uses the fact that even within a linaerly
realized SBGT the
scalar fields can be parametrized nonlinearly \cite{gkko,lezj,clt}
by the point transformation
\be  \Phi \to \frac{v+h}{\sqrt{2}}U \label{trafo}\ee
(with $U$ and $\Phi$ given by \eref{phi}).
The Lagrangian of the Higgs model
in which the scalar sector is nonlinearly realized,
\be \L{}^{H,S} \equiv \L{}^{H}\big|_{\cxiii}, \label{leffhs}\ee
describes a Stueckelberg model with one additional
physical scalar $h$.
Thus (remembering the last paragraph of section 2),
the  results of the previous two
sections can be used to quantize $\L{}^{H,S}$;
the generating functional takes the Lagrangian form \eref{lpi}
with the quantized Lagrangian
\be \L{quant}= \L{U}^{H}\equiv \L{}^{H,S}\big|_{\cxiv} =
\L{}^{H}\big|_{\cxiv}\label{lu}.\ee
It is now easy to establish the
equivalence between $\L{}^{H}$ and $\L{}^{H,S}$
since a point transformation (i.e., a transformation which
does not involve
derivatives) like \eref{trafo} becomes a canonical
transformation within
the Hamiltonian formalism, i.e.\ the Hamiltonians
and also the
constraints corresponding to $\L{}^{H}$ and $\L{}^{H,S}$
are related by
canonical transformations\footnote{For the Hamiltonian
and the primary
constraints this statement is obvious and the secondary
constraints are obtained from the Poisson brackets
\eref{eom} which are
invariant under canonical transformations.}.
Thus, the physical systems described by both
Lagrangians are equivalent on the
Hamiltonian level. $\L{U}^H$ becomes the U-gauge of $\L{}^H$;
i.e., also for a
linearly realized Higgs model the U-gauge is obtained naively
by removing all
unphysical scalar fields.

Due to the invariance of the Hamiltonian PI under
canonical transformations
\cite{fad}, the generating
funtional obtained when quantizing the
linear Lagrangian $\L{}^H$  also has the
form
$\eref{lpi}$ with $\eref{lu}$, which is again identical
(apart from $\df$ terms) to the result of the
Faddeev--Popov procedure if the (U-gauge)
g.f.\ conditions \eref{pgf} are applied\footnote{Remember
footnote \ref{ug}.}
\cite{gkko}. As in the
previous section, this result can be generalized to
any other gauge. This
completes the proof of Matthews's theorem for any
effective Lagrangian, which
fulfils the requirements listed at the beginning of section~2.

The treatment of this
section shows that the Stueckelberg formalism, which was
originally introduced
in order to construct Higgs-less SBGTs \cite{stue,bulo}, also
represents a powerful tool when dealing with Higgs models
\cite{gkko,clt}.

\section{The Quartically Divergent Higgs Self-Interaction}
\typeout{Section 5}
\eqnew
In all previous sections, I have neglected the quartically
divergent $\df$ terms.
In this section I will quantize the SU(2) Higgs model
(without effective
non--Yang--Mills vector boson self interactions)
{\em thereby taking into
account the $\df$ terms\/} to derive
the well known quartically divergent
Higgs self
interaction \cite{gkko,leya,lezj,giro},
which serves as a simple example for such a term.

{}From the discussion of the previous two sections it is
clear that it makes
no difference to quantize the gauge invariant Lagrangian
of a SBGT or the
corresponding U-gauge
Lagrangian\footnote{When establishing
this equivalence, no $\df$ terms have
been neglected, thus, even concerning the quartically
divergent extra terms,
quantization of both Lagrangians yields the
same result.} which is obtained
by setting $\ph_a=0$. Thus, for simplicity, I start
from the U-gauge Lagrangian
\be\L{}= -\frac{1}{4}F^{\mu\nu}_aF_{\mu\nu}^a+
\frac{1}{2}\d_\mu h\d^\mu h+
\frac{1}{8}g^2(v+h)^2A^\mu_aA_\mu^a-V(h,\ph_a=0).\label{lsu2}\ee
The momenta are given by
\bea \pi_0^a&=&=\frac{\d\L{}}{\d \da^0_a}=0,\label{pnh}\\
\pi_i^a&=&\frac{\d\L{}}{\d \da^i_a}=F_{i0}^a=
\dot{A}^i_a+\d_iA_0^a+gf_{abc}A^b_iA^c_0,
\label{pih}\\
\pi_h&=&\frac{\d\L{}}{\d \dot{h}}=\dot{h},\label{ph}\eea
and the Hamiltonian is
\bea \H{}&=&\pi_\mu^a\da^\mu_a+\pi_h\dot{h}-\L{}\nn
&=&\frac{1}{2}\pi_i^a\pi_i^a+\frac{1}{2}\pi_h^2-
\pi_i^a\d_iA^0_a-gf_{abc}\pi^a_i
A^b_iA^c_0+\frac{1}{4}F_{ij}^aF_{ij}^a+
\frac{1}{2}(\d_ih)(\d_ih)\nn&&-
\frac{1}{8}g^2(v+h)^2 (A_0^aA_0^a-
A_i^aA_i^a)+V(h,\ph_a=0).\label{hh}\eea
The constraints turn out to be
\bea\phi_1^a&=&\pi_0^a=0,\label{pch}\\
\phi_2^a&=&\d_i\pi_i^a-gf_{abc}\pi^b_i A^c_i-
\frac{1}{4}g^2(v+h)^2 A_0^a=0.
\label{sch}\eea
The Poisson bracket of the primary and the
secondary constraints is given by
\be \{\phi_1^a(x),\phi_2^b(x)\}=
\frac{1}{4}g^2(v+h)^2\delta^{ab}\delta^4(x-y).
\label{pbh}\ee
The constraints are second class.

To quantize this, one starts from the Hamiltonian PI \eref{hpi},
integrates out
the $\pi_0^a$ to eliminate $\delta(\phi_1^a)$, uses
\eref{exp} to rewrite
$\delta(\phi_2^a)$, performs the substitution \eref{subs} and
rewrites the deteminant using
\eref{det} and \eref{pbh}. The generating functional becomes
\bea Z[J]&=&\int\prod_{\mu,a}\D A_\mu^a\D h
\prod_{i,a}\D\pi_i^a\D\pi_h\prod_a\D\lambda^a\,
\exp\bigg\{ i\int\dx\bigg[-\frac{1}{2}\pi_i^a\pi_i^a
-\frac{1}{2}\pi_h^2+  \pi_i^a
F_{i0}^a+\pi_h\dot{h}\nn&&-\frac{1}{4}F_{ij}^a
F_{ij}^a-\frac{1}{2}(\d_i h)(\d_i h)+
\frac{1}{8}g^2(v+h)^2( A_0^aA_0^a-\lambda^a\lambda^a
-A_i^aA_i^a)\nn&&-V(h,\ph_a=0)+J_\mu^aA^\mu_a+J_h h\bigg]
\bigg\}\,{\rm Det^3}\left(\frac{1}{4}g^2(v+h)^2\delta^4(x-y)
\right).\eea
Now one can perform the Gaussian intergrations over
$\pi_i^a$, $\pi_h$ and $\lambda^a$. Intergrating out
$\lambda^a$ yields an extra factor
${\rm Det^{-\frac{3}{2}}}\left(\frac{1}{4}g^2(v+h)^2
\delta^4(x-y)\right)$.
One finds \be Z[J]=\int\prod_{\mu,a}\D A_\mu^a\D h\,
\exp\left\{i\int\dx[\L{}+
J_\mu^aA^\mu_a+J_h h]\right\}\Det\left(
\frac{1}{8}g^3(v+h)^3\delta^4(x-y)
\right).\label{aa}\ee
Using \eref{expdet} to exponentiate the determinant,
$Z[J]$ becomes a
Lagrangian PI
\eref{lpi} with the quantized Lagrangian
\be \L{quant}=\L{}-3i\delta^4(0)\ln\left(1+\frac{h}{v}\right)
=\L{}-3i\delta^4(0)\ln\left(1+\frac{g}{2M}h\right)\label{extra}\ee
(after dropping a constant). Thus, the quantized Lagrangian
contains, in addition to the primordial Lagrangian,
an extra quartically divergent Higgs self-interaction term.

Alternatively, the
determinant in \eref{aa} can be exponentiated by
introducing Grassmann
variables, which yields the ghost term
\be\L{ghost}=-M\eta^\ast_a\eta_a-\frac{g}{2}\eta^\ast_a\eta_a h.
\label{ghost}\ee
The ghost fields are static due to the absence of a kinetic term.
Thus, all ghost loops are quartically divergent.
In \cite{gkko} it has been
shown that the ghost loops following from \eref{ghost} yield the
same contribution to the $S$-Matrix elements as the $\df$ term in
\eref{extra} and thus $\L{ghost}$ can be replaced by this term.

For the renormalizable Lagrangian \eref{lsu2}, however,
the quartic
divergences from the extra
term in \eref{extra} cancel against other
quartically
divergent Higgs self interactions arising from vector
boson loops \cite{apqu}.
Thus, in this case it is completely
justified to neglect the quartic divergences
altogether as in \cite{bedu}.

\section{Summary}
\typeout{Section 6}
\eqnew
The quantization of Lagrangians containing arbitrary interactions of
massive vector fields (that do not depend on higher
order derivatives) within
the Hamiltonian PI formalism yields the following results:
\begin{itemize}
\item The generating functional corresponding to
an effective Lagrangian
without gauge freedom is a simple Lagrangian
PI with the quantized Lagrangian being identical
to the primordial one
(apart from $\ep^2$ and $\df$ terms). Thus, the Feynman rules
follow directly from the various terms in
the effective Lagrangian.
\item (Linearly or nonlinearly realized) SBGTs containing
effective vector boson interactions, which are embedded
in a gauge invariant framework,
can be quantized within the (Lagrangian)
Faddeev--Popov PI formalism.
\item The U-gauge of such an effective
SBGT is obtained by removing all unphysical pseudo-Goldstone fields.
\item Lagrangians related by a Stueckelberg
transformation describe equivalent
physical sytems.
\item Using the Stueckelberg formalism, one can
rewrite each effective Lagrangian as a
(nonlinearely realized) SBGT and extend it,
by introduction of (a) physical
scalar(s), to a (linearly realized) Higgs model.
\end{itemize}
These statements seem to be obvious from the
naive Lagrangian point of view.
However, one has to go through the more elaborate
Hamiltonian treatment of this
paper to derive them correctly.

\section*{Acknowledgement}
I am very grateful to Reinhart K\"ogerler for his encouragement
and advice, for many stimulating discussions and for reading the
manuscript of this paper.

\end{document}